\documentclass[aps,pra,preprint,amsfonts,amsmath,floatfix,showpacs,footnoteintext]{revtex4-1}
 \usepackage[final]{graphicx}
\begin{document}
\title{An improved proximity force approximation for electrostatics}
\author{C\'esar D. Fosco$^{1,2}$}
\author{Fernando C. Lombardo$^3$}
\author{Francisco D. Mazzitelli$^{1,3}$}

\affiliation{ $^1$ Centro At\'omico Bariloche,
Comisi\'on Nacional de Energ\'\i a At\'omica,
R8402AGP Bariloche, Argentina}
\affiliation{ $^2$ Instituto Balseiro,
Universidad Nacional de Cuyo,
R8402AGP Bariloche, Argentina}
\affiliation{ $^3$ Departamento de F\'\i sica {\it Juan Jos\'e
Giambiagi}, FCEyN UBA, Facultad de Ciencias Exactas y Naturales,
Ciudad Universitaria, Pabell\' on I, 1428 Buenos Aires, Argentina - IFIBA}

\date{today}

\begin{abstract} 
A quite straightforward approximation for the electrostatic interaction
between two perfectly conducting surfaces suggests itself when the distance
between them is much smaller than the characteristic lengths associated to
their shapes. Indeed, in the so called ``proximity force approximation''
the electrostatic force is evaluated by first dividing each surface into a
set of small flat patches, and then adding up the forces due two opposite
pairs, the contribution of which are approximated as due to pairs of
parallel planes.  This approximation has been widely and successfully
applied to different contexts, ranging from nuclear physics to Casimir
effect calculations.  We present here an improvement on this approximation,
based on a derivative expansion for the electrostatic energy contained
between the surfaces. The results obtained could be useful to discuss the
geometric dependence of the electrostatic force, and also as a convenient
benchmark for numerical analyses of the tip-sample electrostatic
interaction in atomic force microscopes.

\vskip 2cm

\noindent
{Corresponding author: F.D. Mazzitelli. EMAIL: fdmazzi@cab.cnea.gov.ar; TEL: 54 2944 445151 Ext39;
FAX 54 2944 445299}

\end{abstract}

\maketitle

\section{Introduction}\label{sec:intro}
A standard problem in electromagnetism is to compute the electrostatic
force between conducting bodies, or its close relative: the calculation of
the capacity of a system of conductors.  The simplest example is the case
of parallel plates separated by a distance much smaller than the
characteristic size of the plates, which are held at a fixed electrostatic
potential difference.  Albeit not as straightforwardly as in that example,
some other systems admit analytical exact solutions also; indeed, that is
the case for two eccentric cylinders, for a cylinder in front of a plane,
and also for a sphere in front of a plane. 

This problem is also of considerable practical relevance in Electrostatic
Force Microscopy (EFM) and its variants~\cite{reviewAFM}, which are based
on the interaction between biased atomic force microscopy (AFM) tips and a
sample.  The same applies to the experimental determination of Casimir or
gravitational forces between conducting bodies, as residual charges or
potentials produce unwanted forces that must be subtracted  in order to
determine the sought-after force~\cite{bookCasimir}. 

Of course the electrostatic force between bodies of arbitrary shape can, in
principle, be computed by solving numerically the Laplace equation with
adequate boundary conditions. However, analytic or semi-analytic methods
are always welcome, as ways to improve the understanding of the geometric
dependence of the force, and also to be used as simple benchmarks of fully
numerical computations. For instance, in the context of EFM/AFM, analytical
models for the tips have been developed, and additional exactly
solvable models have been found, like the case of an hyperboloid in front of a
plane~\cite{reviewtips}. Generalized image-charge methods have also been
proposed~\cite{GIM}, in which the tip and the sample are replaced by a set of
fictitious charges, whose intensity and positions are found numerically.

An interesting analytical approach was introduced by B.~Derjaguin in 1934,
who developed an approximate method to compute the Van der Waals force
between macroscopic bodies assumed to be close to each
other~\cite{Derjaguin,surfaceforces}. The approximation assumes that the
surfaces of the bodies can be replaced by a set of flat patches and that at
short distances the dominant contributions correspond to pairs of  patches
(one on each surface) which are closest to each other. Moreover, the
interaction is supposed to be additive. In this way, it is possible to
compute the force between gently curved surfaces from the knowledge of the
interaction energy  for flat surfaces, as long as the radii  of curvature of
the surfaces are much larger than the minimum distance between bodies and
when the surface normals in opposite patches are almost parallel.  Later
on, the same idea was applied to nuclear physics, under the name of
Proximity Force Approximation (PFA) or Proximity Force Theorem, in order to
compute the interaction between nuclei~\cite{Blocki,reviewnuclear}.  It has
also been widely used to compute Casimir (or retarded Van der Waals) forces
between neutral macroscopic objects \cite{bookCasimir}.  It should be clear
that, {\em mutatis mutandis}, the Derjaguin approximation and its ulterior
developments can also be applied to the analysis of electrostatic forces,
at least between gently curved conducting surfaces.

In spite of the simplicity and usefulness of the PFA, for many years there
was a stumbling block to its progress, since methods to asses its
reliability, or to compute the next to leading order (NTLO) correction were
lacking.  This situation  left, as the only alternative of assessing the
PFA reliability, its comparison with the rather few examples for which an
exact result was available.  

In an attempt to improve that situation, we have recently
shown~\cite{depfa}, in the context of Casimir physics, that the PFA can be
considered as the first term, in an expansion in derivatives of the
surfaces shapes, of the interaction energy.  In this way, it is now
possible to improve the PFA by computing the NTLO corrections. 

In this paper we carry out this idea
in the context of electrostatics,
as already suggested in~\cite{depfa}.
 Our aim is twofold: on the one hand, to show the potential usefulness of the improved PFA in computing  electrostatic forces.
On the other hand,  we believe that this attempt to present the improved PFA in a simpler context, may help one to gain intuition about its 
applicability in more complex scenarios.

\section{Proximity force approximation} \label{sec:pfa}
Let us assume, for simplicity's sake, that the system consists of a gently
varying surface in front of a plane (both perfect conductors). The plane is
at $z=0$, and we assume it to be grounded. The curved surface is described
by a single function $z=\psi(x,y)$, and is assumed to be at a constant
electrostatic potential $V$.  We shall use the notation $x_\parallel =
(x,y)$. The electrostatic energy contained between surfaces is then given
by:
\begin{equation} 
 U=\frac{\epsilon_0}{2}\int d^2x_\parallel\int_0^{\psi(x_\parallel)} dz\,
 \vert \mathbf E\vert^2\, .
 \end{equation} 
The force between the conductors can be obtained by computing the
variation of $U$ under a rigid displacement of one of the surfaces, while
the capacity of the system is given by $C=2U/V^2$. 

In the PFA, this system is replaced by a set of parallel plates  (see
Fig.1). The intensity of the electric field between parallel plates
separated by a distance $\psi$ has the $z$-independent value $V/\psi$.
Therefore, the electrostatic energy is approximated by 
\begin{equation}
	U_{\rm PFA} =\frac{\epsilon_0 V^2}{2}\int d^2x_\parallel\, \frac{1}{\psi}\, .
\label{UPFA}
\end{equation}
When computing Van der Waals, nuclear or Casimir forces, the rather
involved nature of the interaction considered can render the interpretation
of the approximations involved in the PFA somewhat obscure.  Here, on the
contrary, the physical assumption is quite clear: the electric field
between opposite patches is regarded as constant and directed along the
line that joins both patches.

Eq.(\ref{UPFA}) can be used to estimate the electrostatic energy in many
interesting situations. To illustrate this point,  let us consider a
cylinder of length $L$ and radius $R$ in front of a plane, and let us
denote by $a$ the minimum distance between surfaces. The function $\psi$,
that describes the part of the cylindrical surface which is closer to the
plane, reads: 
\begin{equation}
\psi(x)=a+R\left(1-\sqrt{1-\frac{x^2}{R^2}}\right)\, ,
\label{psicyl}
\end{equation} 
with $-x_M<x<x_M<R$ in order to cover the part of the cylinder which is
closer to the plane. It is reasonable to assume that $x_M/R=O(1)<1$, and under this assumption the final result will not depend on $x_M$, as long as $R\gg a$, where  the PFA
is expected to give an accurate value of the electrostatic energy. 
Inserting Eq.(\ref{psicyl}) into Eq.(\ref{UPFA}), the integral can be computed
in terms  of elementary functions. Expanding that result for $a\ll R$ we obtain
\begin{equation}
U_{\rm PFA}^{\rm cp} \approx\frac{\epsilon_0V^2L\pi}{\sqrt{2}}\sqrt{\frac{R}{a}}\, ,
\label{PFAcp}
\end{equation}
which is independent of $x_M$. Note that, when the cylinder is close to the
plane, the electrostatic force is proportional to $a^{-3/2}$.

\begin{figure}
\centering
\includegraphics[width=8cm , angle=0]{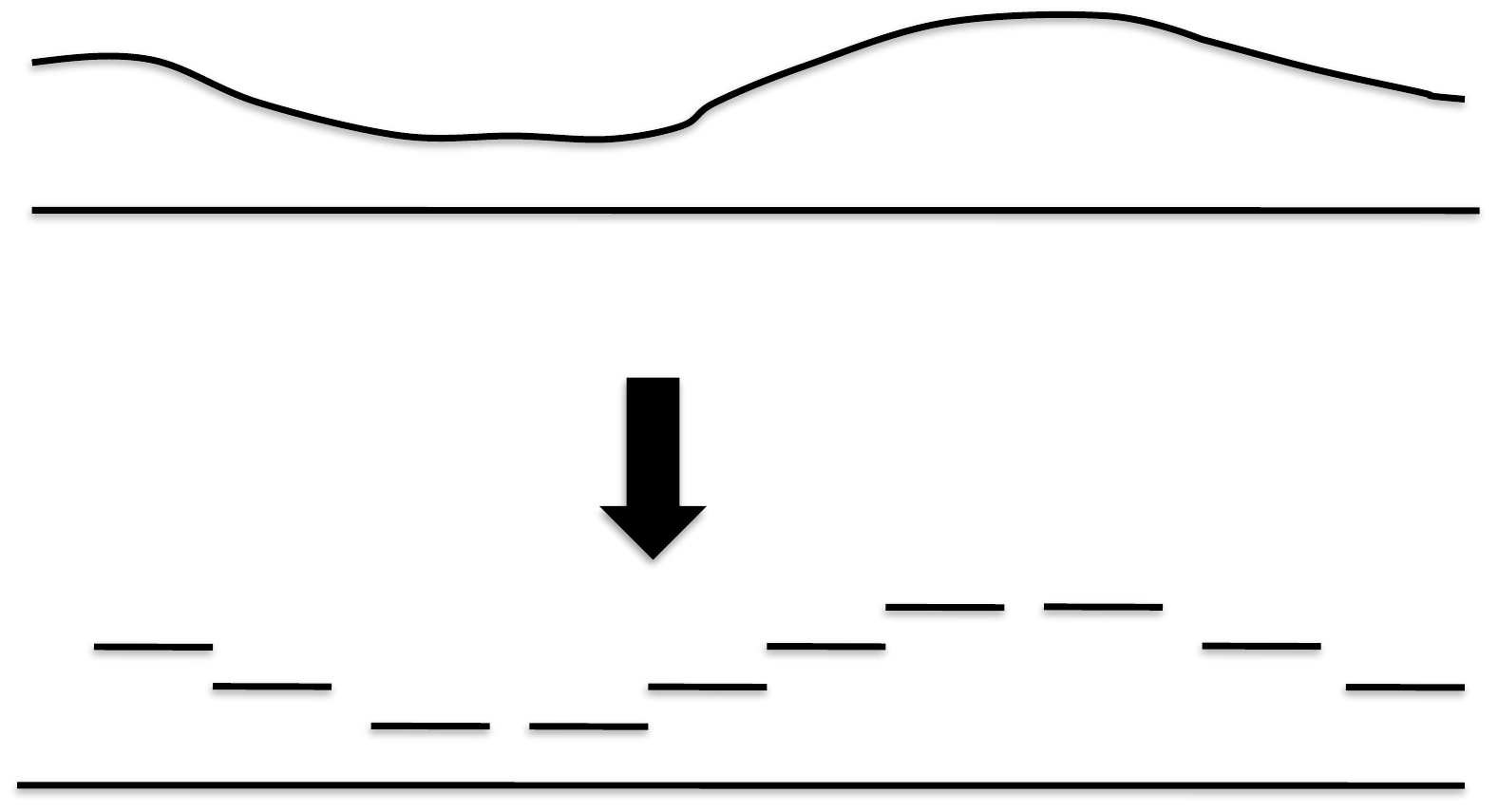}
\caption{\small{In the Derjaguin approximation, the interaction between a gently curved surface and a plane is approximated by
that of  a set of parallel plates. For each pair of parallel plates, border effects  are ignored.}} \label{FigPFA}
\end{figure}

For this geometry, the exact electrostatic interaction energy is
known~\cite{Landau}:
\begin{equation}
 U^{\rm cp} =  \frac{\pi L \epsilon_0 V^2}{{\rm arccosh}\left(1 + \frac{a}{R}\right)}.
\end{equation}
In the limit of $a/R\ll 1$, this exact electrostatic energy indeed reduces
to the PFA result given in Eq.(\ref{PFAcp}). For later use, and to estimate
the relevance of the corrections to the PFA, we expand the
exact result including the NTLO in $a/R$:
\begin{equation}
 U^{\rm cp} \approx \frac{\epsilon_0V^2L\pi}{\sqrt{2}}\sqrt{\frac{R}{a}} \left(1 + \frac{1}{12} \frac{a}{R}\right )
 \equiv U^{\rm cp}_{\rm NTLO}.
 \label{cpntlo}
 \end{equation}

In Fig.2  we plot the ratios $U^{\rm cp}/U_{\rm PFA}^{\rm cp}$ and  $U^{\rm
cp}/U^{\rm cp}_{\rm NTLO}$ as a function of $a/R$. We see that the NTLO produces
a noticeable  improvement of the  PFA. 
We will now compute such improvement for an arbitrary geometry, of which
the exact electrostatic energy in not necessarily known. 
\begin{figure}
\centering
\includegraphics[width=8cm , angle=0]{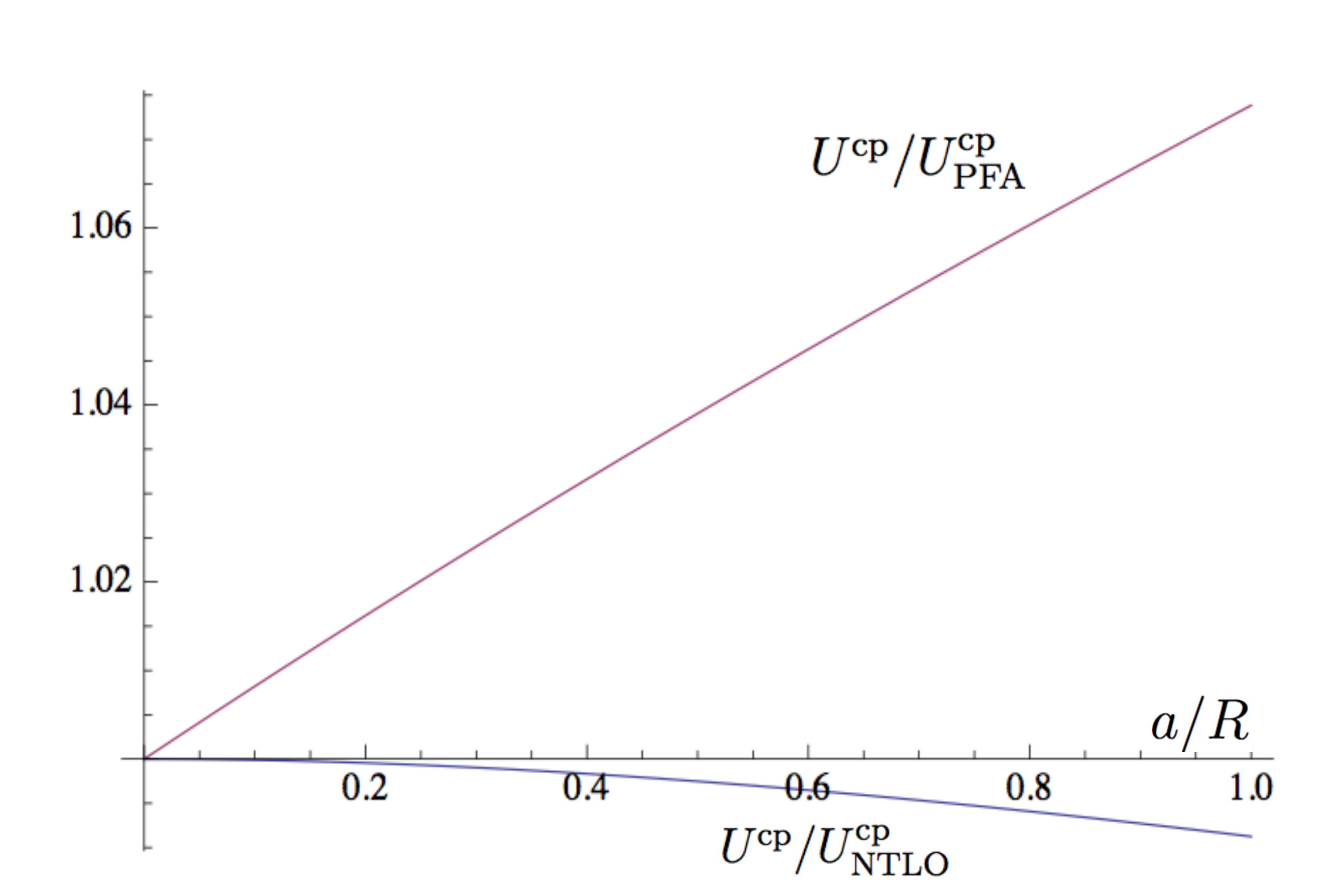}
\caption{\small{Ratio of the exact energy and PFA (upper curve) for the cylinder in
front of a plane, as a function of $a/R$.
For the same geometry, the lower curve corresponds to the ratio of the exact
energy and the approximation that includes the NTLO correction. The inclusion of the NTLO 
improves notably the approximation.}} 
\label{Figcp}
\end{figure}

\section{Improving the PFA: derivative expansion}  \label{sec:de}
In order to improve the PFA, we shall use the fact that the electrostatic
energy may be thought of as a functional of the shape of the surface.  This
functional will be, in general, a non-local object, which becomes local
when the surfaces are sufficiently close and parallel to each other.  To
interpolate between those two situations, we shall therefore assume that
the electrostatic energy can be expanded in derivatives of $\psi$.  Indeed,
one can think of the condition $\vert\nabla\psi\vert\ll 1$, as the
existence of a small, dimensionless parameter. This translates into the
fact that the upper surface is almost parallel to the plane. 

Including terms with up to two derivatives, the electrostatic energy
has to be of the form:
\begin{equation} U_{\rm DE}\simeq\int
	d^2x_\parallel\left[V_{\rm eff}(\psi)+Z(\psi)\vert\nabla\psi\vert^2\right]\,
	, \end{equation} 
for some functions $V_{\rm eff}$ and $Z$ \cite{notation}.
The result must be proportional to $\epsilon_0 V^2$, and must reproduce $U_{\rm PFA}$
for constant $\psi$. Moreover, as there are no other dimensional quantities beyond $\psi$,
dimensional analysis imply that both functions $V_{\rm eff}$ and $Z$ must be proportional to $\psi^{-1}$:
\begin{equation}
U_{\rm DE}\simeq \frac{\epsilon_0V^2}{2}\int d^2x_\parallel\,  \frac{1}{\psi}(1+\beta_{\rm em}\vert\nabla\psi\vert^2)\, ,
\label{PFAimproved}
\end{equation}
where $\beta_{\rm em}$ is the only (numerical) coefficient to be determined.
Note that this coefficient is {\em independent of the nature of the surface
being considered}; therefore, it may be obtained once and for all from its
evaluation in a single case. Indeed, in order to obtain  the coefficient $\beta_{\rm em}$, the
simplest approach, when an exact solution of the problem is available, 
would be to read its value by expanding the exact solution.  For instance,
for the particular configuration of a cylinder in front of a plane,  one can insert
Eq.(\ref{psicyl}) into Eq.(\ref{PFAimproved}), perform the integrals, and
expand the result in powers of $a/R$. Comparing this with the expansion of the
exact result given in Eq.(\ref{cpntlo}), it is straightforward to show that
$\beta_{\rm em}=1/3$.  
Of course the same result could have been obtained from any other particular
example. 

Nevertheless, the coefficient $\beta_{\rm em}$ can also be computed from
first principles, as described in the next section.  \section{Solving the
Laplace equation}\label{sec:laplace} It is perhaps more satisfying to prove
that $\beta_{\rm em}=1/3$ without assuming a particular shape for the
curved surface, but rather for a rather general class of surfaces. To that
end it will be sufficient to derive it by assuming that $\psi(x_\parallel)=
a + \eta(x_\parallel)$ where $a$ is the mean value of $\psi$, and
$\vert\eta(x_\parallel)\vert\ll 1$.  For this class of surfaces, the
derivative expansion of the electrostatic energy reads
\begin{equation}
U_{\rm DE}\simeq \frac{\epsilon_0V^2}{2a}\int d^2x_\parallel \left[1+\left(\frac{\eta}{a}\right)^2+\beta_{\rm em}\vert\nabla\eta\vert^2
+O(\eta^3)\right]\, ,
\label{form}
\end{equation}
and therefore he coefficient $\beta_{\rm em}$ can be read from 
the quadratic dependence of $U_{\rm DE}$ on the derivatives of $\eta$. 
 
The electric field is given by $\mathbf E=-\nabla\phi$, where $\phi$ is
the electrostatic potential, which satisfies the Laplace equation with the
appropriate boundary conditions on the surfaces:
\begin{equation}
\nabla^2\phi =0, \quad \phi_{z=0}=0, \quad \phi_{z=\psi}=V.
\end{equation}

To proceed, we follow the approach of~\cite{Ford}, to  trade a simple
boundary condition on a complex surface into a complex boundary condition
on a flat surface. We expand the boundary condition on the upper surface as
\begin{equation}
V=\phi(x_\parallel,a+\eta)=\phi(x_\parallel,a)+\eta\partial_z\phi(x_\parallel,a)+\frac{1}{2}\eta^2\partial_z^2\phi(x_\parallel,a) 
+....\, ,
\label{expbc}
\end{equation}
and we expand the electrostatic potential as
\begin{equation}
\phi=\phi_0+\phi_1+\phi_2+\dots\, ,
\label{expphi}
\end{equation}
where $\phi_0$ is the solution for parallel plates separated by a distance $a$, that is $\phi_0=V z/a$, and
$\phi_i$ is assumed to be $O(\eta^i)$.

Inserting Eq.(\ref{expphi}) into Eq.(\ref{expbc}), we see that $\phi_1$ and
$\phi_2$ must satisfy the following boundary conditions on $z=a$:
\begin{eqnarray}
\phi_1(x_\parallel,a)&=&-\frac{V}{a}\eta(x_\parallel)\nonumber\\
\phi_2(x_\parallel,a)&=&-\eta(x_\parallel)\partial_z\phi_1(x_\parallel,a)\, .
\label{approxbc}
\end{eqnarray}
Thus, the initial problem has been converted into a problem of parallel
plates for $\phi_1$ and $\phi_2$, with more involved boundary conditions. 
Up to quadratic order, the electrostatic energy reads
\begin{equation}
U_{\rm DE}\simeq\frac{\epsilon_0}{2}\int d^2 x_\parallel\int_0^{a+\eta}dz\left[
\frac{V^2}{a^2}+\vert\nabla\phi_1\vert^2+2\frac{V}{a}\partial_z\phi_1+2\frac{V}{a}\partial_z\phi_2\right].
\label{Uquad}
\end{equation}

It is easy to check that the last two terms in Eq.(\ref{Uquad}) do cancel:
\begin{equation}
\int d^2 x_\parallel\int_0^{a+\eta}dz\, (\partial_z\phi_1+\partial_z\phi_2)=\int d^2 x_\parallel\, \eta(x_\parallel)=0\, ,
\label{cancel}
\end{equation}
where we have used that $(\phi_1+\phi_2)\vert_{a+\eta}=\eta+o(\eta^2)$ and that the mean value of $\eta$ vanishes.

Therefore the correction to the PFA is given solely by the term proportional to $\vert\nabla\phi_1\vert^2$. The order-$\eta$ solution $\phi_1$ is given by 
\begin{equation}
\phi_1(x_\parallel,z)=-\frac{V}{a}\int\frac{d^2k_\parallel}{(2\pi)^2}e^{ik_\parallel x_\parallel}\tilde
\eta(k_\parallel)\frac{\sinh k z}{\sinh k a}\, .
\label{phi1}
\end{equation}
Indeed, it is simple to check that it satisfies both the Laplace equation
and the correct boundary condition (we are using the notation
$k\equiv\vert k_\parallel\vert$).  The solution for $\phi_2$ is
similar-looking, although as already pointed out its explicit expression will not be necessary in
order to evaluate the second-order energy.

Using the solution Eq.(\ref{phi1}) we obtain:
\begin{equation}
\int d^3x\vert\nabla\phi_1\vert^2=\frac{V^2}{a^2}\int\frac{d^2k_\parallel}{(2\pi)^2}\vert\tilde\eta(k_\parallel)\vert^2
F(k_\parallel), 
\end{equation}
where
\begin{equation}
F(k_\parallel)=\int_0^adz \frac{k^2}{\sinh^2k a}(1+2\sinh^2 k z)\simeq \frac{1}{a}+\frac{a k^2}{3}\, .
\end{equation}
(note that the integration in $z$ runs up to $a$ instead of $a+\eta$ since the integrand is already
of order $\eta^2$).
Therefore
\begin{equation}
\frac{\epsilon_0}{2}\int d^3x\vert\nabla\phi_1\vert^2\simeq\frac{\epsilon_0V^2}{2 a}\int d^2x_\parallel\, 
\left[\left(\frac{\eta^2}{a^2}\right)+
\frac{1}{3}\vert\nabla\eta\vert^2\right]\, ,
\end{equation}
which is of the form of Eq.(\ref{form}), and shows that  $\beta_{\rm em}=1/3$.  
The final
result is then
\begin{equation}
U_{\rm DE}\simeq \frac{\epsilon_0V^2}{2} \int d^2x_\parallel\,  \frac{1}{\psi}\left[1+\frac{1}{3}\vert\nabla\psi\vert^2
\right]\, ,
\label{EE}\end{equation}
as expected from the example of the cylinder in front of a plane.
Eq.(\ref{EE}) is the main result of this paper. The vertical component of
the force on the upper surface can be computed by taking the variation of
the energy under an infinitesimal
translation $\psi\rightarrow\psi + \delta$, that is
\begin{equation}
F_z\simeq\frac{dU_{\rm DE}}{d\delta}\simeq - \frac{\epsilon_0V^2}{2} \int d^2x_\parallel\,  \frac{1}{\psi^2}\left[1+\frac{1}{3}\vert\nabla\psi\vert^2\right]\, .
\label{FE}\end{equation}
Note that, as the conductors are not isolated (they are kept at a fixed potential difference), the force is given
by the derivative of the energy with respect to the position, with a positive sign.

\section{Tip-sample electrostatic interaction}
In this section we use the improved PFA to analyze the problem of the
tip-sample electrostatic force in an AFM. Typical geometries to model the
interaction are  sphere-plane, paraboloid-plane, hyperboloid-plane, and
sphere-ended cone in front of a plane, among others \cite{reviewtips}.  In
all these  cases, it is  possible to obtain the electrostatic force using
the improved PFA. The results will be accurate when the tip is sufficiently
close to the sample. 
\subsection{Sphere-plane}
As a first example, we describe the case of a sphere of radius $R$, in
front of a plane at a minimum distance $a$ ($a\ll R$).  The surface of the
sphere cannot be described by a single valued function $z =
\psi(x_\parallel)$. We will just consider the region of the sphere which is
closer to the plane, and we will see that it still give quantitatively
adequate results even beyond the lowest order.  The function $\psi$ is
\begin{equation}
\psi(\rho) = a + R \left(1 - \sqrt{1 - \frac{\rho^2}{R^2}}\right),
\end{equation}
where we are using polar coordinates ($\rho, \phi$) in the plane ($x,y$).
This function describes the hemisphere when $0\leq \rho \leq R$. The
derivative expansion will be well defined if we restrict the integrations
to the region $0\leq \rho\leq\rho_M<R$.  As in the example of the cylinder in front of a plane,
described in Section II, we will assume that $\rho_M/R=O(1)<1$.  Inserting this expression for
$\psi$ into the derivative expansion for the electrostatic energy
Eq.(\ref{EE}), one can perform explicitly the integrations and obtain an
analytic expression $U_{\rm DE}(\rho_M,a,R)$. Expanding the result for
$a/R\ll 1$ we obtain
\begin{equation}
U_{\rm DE} \simeq U_{\rm PFA}^{\rm sp}+f_1 + \frac{1}{3} \frac{a}{R}\left ( U_{\rm PFA}^{\rm sp}+f_2
\right), 
\label{desp1}
\end{equation}
where the leading contribution is given by 
\begin{equation}
U_{\rm PFA}^{\rm sp} = - \pi V^2\epsilon_0 R
\log(2a/R)\,,
\end{equation}
and $f_{1,2}$ are functions of $\rho_M/R$. Being independent of $a$,  $f_1$
can be discarded if one is interested in computing the force between the
sphere and the plane. The term proportional to $f_2$ gives a next to NTLO
contribution to the force. Therefore we write the final result as 
\begin{equation}
U_{\rm DE} \simeq U_{\rm PFA}^{\rm sp} \left( 1 + \frac{1}{3} \frac{a}{R}\right).
\label{desp}
\end{equation}
Note that, as long as $a\ll R$, the force will not depend on 
$\rho_M$:
\begin{equation}
F_z\simeq\frac{dU_{\rm DE}}{da}=-\frac{\pi V^2\epsilon_0 R}{a}\left(1+\frac{1}{3}\frac{a}{R}\log\left(\frac{a}{R}\right)\right)\, .
\label{Fsp}
\end{equation}
This result could also be obtained directly from Eq.(\ref{FE}), avoiding the 
appearance of the $\rho_M$-dependent constant terms $f_{1,2}$. 
 
As already mentioned, this geometric configuration admits an
exact solution \cite{smythe}; at short distances that exact result reduces
to Eqs.(\ref{desp}) and (\ref{Fsp}), as discussed in Ref. \cite{brasilia} (we include some
details in the Appendix).

Finally, it is worth to remark that at short distances the electrostatic
force for this configuration is proportional to $a^{-1}$, milder than the
cylinder-plane geometry ($a^{-3/2}$) and than the parallel plates
($a^{-2}$). This is of course a consequence of the different effective
areas of interaction in each case, and has its counterpart in Casimir
physics \cite{EPL}.  It is a simple property that could be emphasized  in
elementary discussions of electrostatics. 
\subsection{Paraboloid-plane}
For the case of a paraboloid-like tip, the surface is described by 
\begin{equation}
\psi(\rho) = a + \frac{\rho^2}{2R},
\end{equation}
with $0\leq \rho\leq \rho_{\rm M}<R$. $\kappa = R^{-1}$ is the local curvature at
the bottom of the paraboloid. 
The derivative expansion for the electrostatic interaction energy is
\begin{equation}
U_{\rm DE} \simeq U_{\rm PFA}^{\rm sp} \left( 1 - \frac{2}{3} \frac{a}{R}\right).
\end{equation}
It is not surprising that the leading order coincides with that of the
sphere-plane interaction, since both surfaces have the same curvature at
the point closest to the plane. However, the NTLO corrections are
different, and in fact have opposite signs.  These corrections can
therefore be used as benchmarks of numerical calculations, since the short
distance regime is typically the complex from a computational point of
view. 
\subsection{Hyperboloid-plane}
In this case the surface is described by the function
\begin{equation}
\psi(\rho) = a -R_1+ R_1  \sqrt{1 + \frac{\rho^2}{R_2^2}}, 
\end{equation} 
where the curvature $\kappa$ at the bottom of the 
hyperboloid is $\kappa=R^{-1}$, with $R= R_2^2/R_1$, 
and $R_1/R_2$ is its asymptotic slope. 
 
As in the previous examples, the derivative expansion of the electrostatic energy can be computed
in terms of elementary functions. The result is
\begin{equation}
U_{\rm DE}\simeq U_{\rm PFA}^{\rm sp}\left[1-(\frac{2}{3}-\frac{R_2^2}{R_1^2})\frac{a}{R}\right]\, .
\end{equation}
It is remarkable that sign of the NTLO correction depends on the parameters
of the hyperboloid, and vanishes for the particular asymptotic slope
$R_1/R_2=\sqrt{3/2}$. 
\section{Generalizations}
The results presented so far can be generalized in several directions:
\subsection{Beyond the next to leading order}
It is in principle possible to extend the results in order to include more
derivatives of the shape function. For example, coming back to the case of
a surface in front of a plane, the improved PFA would be, including the
next to NTLO corrections:
\begin{eqnarray}
\label{4derivatives}
U_{\rm DE}&=& \frac{\epsilon_0V^2}{2} \int d^2x_\parallel\,  \frac{1}{\psi}(1+\frac{1}{3}\vert\nabla\psi\vert^2
+\gamma_1\vert\nabla\psi\vert^4+\gamma_2\psi\vert\nabla\psi\vert^2\nabla^2\psi \nonumber \\ &+& \gamma_3\psi^2
\nabla^2\psi\nabla^2\psi+ \gamma_4 \psi^2 \partial_\alpha \partial_\beta \psi \partial_\alpha \partial_\beta \psi + \gamma_5 \psi^3 \nabla^2\nabla^2 \psi),
\end{eqnarray}
where $\gamma_1,\gamma_2,...
\gamma_5$,  are numerical coefficients that can be determined extending the calculation of
Section \ref{sec:laplace} up to $O(\eta^4)$ \cite{f2}.

The structure of the improved PFA is useful to analyze its range of
validity: the NTLO corrections are small when $\vert \nabla \psi\vert \ll
1$ or, in other words, when the curved surface is almost parallel to the
plane. Higher order corrections will be negligible when, in addition to
this condition, the scale of variation of the shape of the surface is much
larger than the local distance between surfaces.
\subsection{Flat surface coated with a dielectric layer}
The improved PFA can be  extended to the case in which the sample (flat
surface) is a perfect conductor coated with  a layer of width $d$ of a
material characterized with a constant permittivity $\epsilon$, as
considered in the context of EFM \cite{layer2011}.  Dimensional analysis
implies that the derivative expansion for the electrostatic energy will be
of the form 
\begin{equation}
U_{\rm DE}\simeq \frac{\epsilon_0 V^2}{2} \int d^2x_\parallel\,  \frac{1}{\psi}\left[\alpha(\frac{\epsilon}{\epsilon_0},\frac{d}{\psi})+
\beta(\frac{\epsilon}{\epsilon_0},\frac{d}{\psi})\vert\nabla\psi\vert^2\right ]\, ,
\label{diel}\end{equation}
for some functions $\alpha$ and $\beta$. We do not dwell here with the
determination of the function $\beta$, but just want to remark that
$\alpha$ can be obtained by elementary considerations for a
parallel-plate capacitor with a dielectric layer: 
\begin{equation}
\alpha(\frac{\epsilon}{\epsilon_0},\frac{d}{\psi})
=\frac{1}{1-\frac{d}{\psi}(1-\frac{\epsilon_0}{\epsilon})}\,.
\end{equation}
\subsection{Two gently curved surfaces}
For the case of the Casimir energy, the result of Ref.\cite{depfa} has been
generalized to the case of two gently curved surfaces in
Ref.\cite{bimonte}. This generalization can also be applied to the
electrostatic case.  Let us assume that the surfaces are described by two
functions $\psi_1(x_\parallel)$ and  $\psi_2(x_\parallel)$
that measure the height of the surfaces with respect to a reference plane $\Sigma$.
The derivative expansion for the electrostatic energy reads, in this case:
\begin{eqnarray}\label{2surfaces}
U_{\rm DE}[\psi_1,\psi_2] &=&\frac{\epsilon_0V^2}{2}\int_{\Sigma} d^2x_\parallel\,\frac{1}{\psi}\, \left[1+\beta_1 
\vert\nabla \psi_1\vert^2 \right.  \nonumber \\
&+&\left.\beta_2\vert \nabla \psi_2\vert^2 +
\beta_\times\nabla \psi_1\cdot\nabla \psi_2 + \beta_{\rm -} \, {\hat {\mathbf z}} \cdot
\nabla \psi_1  \times \nabla \psi_2) + \cdots\right],
\end{eqnarray}
where $\psi=\vert \psi_2-\psi_1\vert$ is the height
difference, and dots denote higher derivative terms.  

As the energy must be invariant under the interchange of $\psi_1$ and
$\psi_2$, $\beta_1=\beta_2$ and $\beta_-=0$. Moreover, in order to
reproduce the previous result we must have  $\beta_1=\beta_2=1/3$. Finally,
the coefficient $\beta_\times$ can be determined using the fact that the
energy should be invariant under a tilt of the reference surface $\Sigma$
\cite{bimonte} or, alternatively, under a simultaneous rotation of both
surfaces. For an infinitesimal rotation of each surface with an angle
$\epsilon$ in the plane $(x,z)$ we have
$\delta\psi_i=\epsilon(x+\psi_i\partial_x\psi_i)$, for $i=1,2$. In order to
simplify the calculation we can assume that, initially, $\psi_1=0$ and that
$\psi_2$ is only a function of $x$. Computing explicity the variation of
$U_{DE}$ to linear order in $\epsilon$ one can show that 
\begin{equation}
\delta U_{\rm DE}=0\Rightarrow \beta_\times=\frac{1}{3}\, ,
\end{equation}
and therefore
\begin{equation}\label{2surfacesfinal}
U_{\rm DE}[\psi_1,\psi_2]=\frac{\epsilon_0V^2}{2}\int_{\Sigma} d^2x_\parallel\,\frac{1}{\psi}\, \left[1+\frac{1}{3}\left(
\vert\nabla \psi_1\vert^2 +\vert \nabla \psi_2\vert^2 +  \nabla \psi_1\cdot\nabla \psi_2\right)
\right].
\end{equation}

While the equality $\beta_1=\beta_2$ is valid for any interaction (as long
as the surfaces are identical), the fact that $\beta_\times =\beta_1$ is
valid only for the particular case of the electrostatic interaction, in
which the leading term is proportional to 
$\psi^{-1}$  (i.e. is not valid for the Casimir energy, that is 
proportional to $\psi^{-3}$ for parallel plates).

Computing the variation of the electrostatic energy Eq.(\ref{2surfaces})
with respect to translations or rotations of one of the surfaces, it is
possible to  obtain  the vertical  and lateral components of the force, as
well as the torque produced by the presence of the other surface.  

\section{Final remarks}\label{sec:conclusions}
The PFA is a very useful approximation introduced by Derjaguin many years
ago to compute Van der Waals forces between macroscopic bodies. The
approximation has been subsequently widely used in rather different
contexts, like nuclear physics and the Casimir effect.
Very recently, the approximation has been interpreted as the leading order
in a derivative expansion of the energy, and this interpretation
allowed to compute, for the first time, the NTLO corrections in a
systematic way. In this paper we have pointed out that the improved PFA
is a powerful tool which may also be used to compute the electrostatic
interaction between conducting surfaces.  
Indeed, with this improved approximation, it is possible to compute the 
electrostatic force with high accuracy when the surfaces are sufficiently 
close to each other. 

Moreover, given the simplicity of the expressions for the energy and the
force, the NTLO could be used to test numerical methods aimed at a
calculation of the force for arbitrary surfaces.  

\section*{Note added:}
While this paper was under review, the revised version of Ref.\cite{bimonte} was published \cite{bimonte2}.
In the new version, the authors mentioned that they have successfully applied the gradient expansion to analyze the electrostatic
interaction between a sphere and a plane, although there are no details of the calculations.  

\section*{Acknowledgements}
This work was supported by ANPCyT, CONICET, UBA and UNCuyo.

\section*{Appendix}

The electrostatic problem of a sphere of radius $R$ in front of a plane can be solved exactly: if the potential difference between the plane and the sphere is $V$,
the electrostatic energy is given by \cite{smythe}
\begin{equation}
U^{\rm sp}=2\pi\epsilon_0 R V^2 \sinh\Gamma\sum_{n\geq 1}\frac{1}{\sinh(n\Gamma)},
\label{uexactsp}
\end{equation}
where $\cosh\Gamma = 1+a/R$. In order to obtain an analytic expression in the limit $\Gamma\rightarrow 0$
we write
\begin{eqnarray}
S\equiv \sum_{n\geq 1}\frac{1}{\sinh(n\Gamma)}&=& S-\int_1^{\infty}\frac{dn}{\sinh(n\Gamma)}+\frac{1}{\Gamma}\ln(\coth\Gamma)\nonumber\\
&=& \frac{\gamma}{\Gamma} + \frac{1}{\Gamma}\ln(\coth\Gamma) + O(\Gamma),
\label{evalsum}
\end{eqnarray}
where $\gamma = 0.5772$. Replacing this expression into Eq.(\ref{uexactsp}) and expanding the result for small $a/R$
we obtain
\begin{equation}
U^{\rm sp} \simeq 
-\pi\epsilon_0 a V^2 \ln (2 a/R)
\left\{1+\frac{1}{3}\frac{a}{R}+O\left (\frac{a/R}{\ln(a/R)}\right )\right\}\, ,
\label{usppfa}\end{equation}
where we omited an irrelevant constant term. This result coincides, up to the NTLO, with the one obtained
using the derivative expansion Eq.(\ref{desp}).


\begin{thebibliography}{bib}
\bibitem{notation} We use the typical notation for the derivative expansion in Quantum Field Theory, in which $V_{\rm eff}$ is the effective potential and $Z$ the wave function renormalization.
\bibitem{f2} One could be tempted to obtain information about the coefficients $\gamma_i$ from the exact result of the cylinder in front of a plane. However, there is a subtle point here, because the next to NTLO corrections for this geometry depend on the quantity $x_M$ introduced in Section \ref{sec:pfa}.
\end{thebibliography}
\end{document}